\documentstyle[aps,pre,epsfig]{revtex}

\begin{document}
\title{Lattice Boltzmann simulations of three-dimensional
single droplet deformation and breakup under simple shear flow}
\author{Haowen Xi \thanks{haowen@bgnet.bgsu.edu} and Comer Duncan \thanks{gcd@chandra.bgsu.edu}\\
Department of Physics and Astronomy \\
Bowling Green State University\\
Bowling Green, OH 43403.}
\date{\today}
\maketitle

\begin{abstract}
We present  three-dimensional numerical simulations
of the classical Taylor experiment on droplet deformation
within a shear flow. We have used the promising 
Lattice-Boltzmann method numerical scheme 
to simulate single droplet deformation and breakup under simple shear flow. 
We first compute the deformation of the droplet and find excellent agreement
with the theoretical prediction. 
We have used the same method to simulate the shear and breakup for
larger values of the shear rate. We find that the Lattice Boltzmann
method used in conjunction with the interface force model of Shan and
Chen results in an excellent treatment of the entire process from
small deformation to breakup into multiple droplets.
Our results
could be extended to study
the rheology of dispersed droplets and 
the dynamics of droplet breakup and coalescence in shear flow.
\end{abstract}


\newpage
\section{Introduction}
The effect of shear flow on droplets of one fluid freely suspended in 
another immiscible fluid is a problem of longstanding interest
\cite{re:gr71} \cite{re:ac83} \cite{re:ra84} \cite{re:tj91} \cite{re:st94}. 
Long ago Taylor \cite{re:ta32} \cite{re:ta34} considered a droplet 
of a Newtonian fluid suspended in the shear 
flow of a second Newtonian fluid. He estimated the 
largest stable droplet radius by balancing the 
surface stresses due to interfacial tension and viscous stress
due to shear.  Taylor found that the deformation of the droplet 
can be expressed in terms of three dimensionless parameters: 
the capillary (or Taylor) number ${\it Ca}= R \eta_{m} \dot{\gamma} /\beta$;
the viscosity ratio $\lambda=\eta_{d} / \eta_{m}$; 
and the density ratio $\kappa=\rho_{d} / \rho_{m}$.
Here $\beta$ is the interfacial tension coefficient,
$\dot{\gamma}$ is the shear rate, and $R$ is the droplet radius.
$\eta_{d}$ and $\eta_{m}$ are the droplet 
and medium viscosity, respectively, and 
$\rho_{d}$ and  $\rho_{m}$ are the droplet and medium density, 
respectively. For two fluids of equal viscosity ($\lambda$=1) and 
equal density (neutrally buoyant, $\kappa $=1) under simple shear flow, 
Taylor obtained a theoretical result for 
{\it small} deformation $D$, where
\begin{equation}
D=(L-B)/(L+B)=(35/32) {\it Ca}.
\end{equation}
Here $L$ and $B$ are the largest and smallest distances from
the droplet surface from its center (the ``major'' and ``minor'' axes).
Thus for a spherical droplet, $D$ is equal to zero.  

From the numerical simulation point of view, the simulation of 
droplet deformation and breakup problem is very difficult. 
The conventional numerical modeling of liquid-liquid system, which 
involves solving hydrodynamic partial differential 
equations, has seen only limited success
\cite{re:st94}. The equations of motion of flow field 
must be solved both inside and outside the droplet, 
with the  appropriate boundary condition applied 
on the interface between the interior and exterior of the droplet. 
However, the shape of the droplet is 
not known {\it a priori}, and must be determined as part of 
the solution. Because of these complications, 
there have not been many successful three-dimensional
numerical studies of droplet deformation and breakup. 
The case of many drops under shear flow where {\it both}
breakup and coalescence (resulting from the collisions of droplets)
effects are taken into account 
is largely unexplored \cite{re:st94}.
In this paper, we report the use  of
an alternative and promising numerical scheme  called 
the Lattice-Boltzmann method (LBM) \cite{re:be92} 
\cite{re:do89}  \cite{re:ro94} \cite{re:shi97}
to simulate the deformation
of droplets under shear.  In recent years there have
been a growing number of successful applications of LBM to a variety of physical systems.
The basic idea of the LBM is to construct 
simplified kinetic models that
incorporate the essential physics of microscopic 
dynamics so that the macroscopic averaged properties obey
the desired macroscopic Navier-Stokes equations.
One of the great advantages of the LBM is 
that the information about the phase boundary (e.g. 
interface boundary between droplets and the exterior medium), the
droplet size and shape, and the flow field 
can all automatically arise from the solutions.
The LBM scheme has been shown  to be particularly successful in multiphase 
flow dynamics \cite{re:gu91} \cite{re:gr93} \cite{re:sh95}
and flow in systems with complex boundaries \cite{re:be92} 
 \cite{re:shi97}. However, there have been only few quantitative studies
reported of three-dimensional multiphase flow problems using the LBM. 
In this paper, we will present the first 
three-dimensional numerical study of droplet deformation 
using LBM with quantitative comparison with
theoretical results in the small deformation regime. An early study 
\cite{re:ha96} of droplet deformation was restricted to
two-dimensional space with no quantitative comparison between the numerical
simulations and the theoretical results for droplet deformation.  The
comparison between the simulations of three dimensional droplet shear
with the theoretical results of Taylor represents an important step in
the validation of the applicability of the LBM method to such physical
systems.

The mechanical dispersion of immiscible droplets is of importance 
both in nature and in many industrial applications
\cite{re:gr71} \cite{re:ac83} \cite{re:ra84}.  
The mixing process under
shear flow is usually divided into three stages: (i) stretching 
and deformation of liquid droplets, (ii) breakup of these droplets
, and (iii) coalescence of the resulting droplets 
upon collision.  The basic process of deformation of a liquid droplet, 
immersed in the given flow field of a second (immiscible) liquid 
is governed by the capillary number, which is  the ratio of the deforming shear stress  
 applied externally and the shape-conserving interfacial
 tension.  One good example of the dispersion of droplets is the 
blending of molten
polymer systems \cite{re:xi96}. Because nearly all chemically different
polymers are immiscible, the effective mixing of immiscible 
polymers is a ubiquitous industrial goal. The usual objective is to 
produce a fine dispersion of sub-micron-sized
particles of one polymer in a matrix of another polymer, with the goal to produce a  
composite system with improved physical properties. Thus, the rheology of 
the dispersion of droplets in various shear flows at low Reynolds numbers 
is of both practical and fundamental interest, and has received considerable 
attention over the past sixty years starting with early
work by G. I. Taylor in 1934. For recent references 
see \cite{re:tj91} \cite{re:st94}. 

This paper is organized as follows:
In section II we present a brief discussion 
of multicomponent LBM and outline our numerical techniques. In section III, we 
present quantitative numerical results for two studies. First, we make a comparison
between the simulation of small deformation droplet with the classical Taylor
theory. Secondly, we report on simulations with a larger shear rate in which
the initial droplet is deformed, then sheared to breakup. We conclude with a 
discussion of some of the interesting
problems which are opened up given the utility of the 
three-dimensional LBM to systems
of droplets under various shear flow conditions.

\section{Numerical Model}
In this section, we present a brief description of
the LBM for modeling multicomponent immiscible fluids 
developed by Shan and Chen (SC) \cite{re:sc93} 
\cite{re:sc94}. Denote by $n^{\sigma}_{a}({\bf{x}},t)$  
the number density  of $\sigma$th particles at spatial point ${\bf{x}}$ and time $t$ for
the fluid ($\sigma=1,2$) with velocity ${\bf{e}}_{a}$.
Here $a=0, ... , b$, and  where $b$ is the number of velocity 
directions on a three-dimensional lattice (in D3Q19 lattice model, $b=18$) 
\cite{re:qi92}.
The LBM for the particle distribution 
function $n^{\sigma}_{a}({\bf{x}},t)$ can be written as follows:
\begin{equation}
n^{\sigma}_{a}({\bf{x}}+{\bf{e}}_{a}, t+1)-n^{\sigma}_{a}({\bf{x}},t)
=-\frac{1}{\tau^{\sigma}} 
[n^{\sigma}_{a}({\bf{x}},t)-n^{\sigma(eq)}_{a}({\bf{x}},t)]
\end{equation}
where $n^{\sigma(eq)}_{a}({\bf{x}},t)$ is the local equilibrium 
distribution function which  depends on the microscopic velocity $\bf{e}_{a}$ and 
the macroscopic density $n_{a}$ , velocity $\bf{u}$, 
and $\tau^{\sigma}$ is the relaxation time 
for species $\sigma$ and controls the rate of approach to equilibrium for that species.
The Galilean-invariant three-dimensional D3Q19 lattice model equilibrium distribution 
function can be represented as:
\begin{equation}
n_{a}^{\sigma(eq)}({\bf{x}},t)=
\frac{1}{3} n^{\sigma}({\bf{x}},t) [1-\frac{3}{2} {\bf{u}} \cdot {\bf{u}}],
|{\bf{e}}_{a}|^{2}=0
\end{equation}
\begin{equation}
n_{a}^{\sigma(eq)}({\bf{x}},t)=
\frac{1}{18} n^{\sigma}({\bf{x}},t) [1+3({\bf{e}}_{a}\cdot {\bf{u}})
+\frac{9}{2}({\bf{e}}_{a}\cdot {\bf{u}})^{2} -
\frac{3}{2}{\bf{u}} \cdot {\bf{u}}],
|{\bf{e}}_{a}|^{2}=1
\end{equation}
\begin{equation}
n_{a}^{\sigma(eq)}({\bf{x}},t)=
\frac{1}{36} n^{\sigma}({\bf{x}},t) [1+3({\bf{e}}_{a}\cdot {\bf{u}})
+\frac{9}{2}({\bf{e}}_{a}\cdot {\bf{u}})^{2} -
\frac{3}{2}{\bf{u}} \cdot {\bf{u}}],
|{\bf{e}}_{a}|^{2}=2
\end{equation}
where the macroscopic density $n^{\sigma}({\bf{x}},t)$ and velocity
for each fluid component $\sigma$ are defined as
\begin{equation}
n^{\sigma}({\bf{x}},t)=\sum_{a} n^{\sigma}_{a}({\bf{x}},t)
\end{equation}
and 
\begin{equation}
{\bf{u}}({\bf{x}},t)
=\frac{\sum_{\sigma} m^{\sigma} \sum_{a} n^{\sigma}_{a} {\bf{e}}_{a} 
          /\tau^{\sigma}}
              {\sum_{\sigma} m^{\sigma} \sum_{a} n^{\sigma}_{a} 
          /\tau^{\sigma}}
+ \frac{\tau^{\sigma}}{\sum_{a} m^{\sigma} n_{a}^{\sigma}({\bf{x}},t)}
\frac{d{\bf{p}}^{\sigma}}{dt}.
\end{equation}
where $m^{\sigma}$ is the mass of the $\sigma$th component.
Note that for the $\sigma$th component the macroscopic mass density and momentum
density are defined to be $\rho_{\sigma} = m^{\sigma} n^{\sigma} $ and 
$\rho_{\sigma} {\bf{u_{\sigma}}} =  m^{\sigma} \sum_{a} n^{\sigma}_{a} {\bf{e}}_{a} $
respectively.
The second term in the above equation represents the interaction between the
two fluid components. In order to model surface tension forces,
SC introduced an 
interaction potential $V({\bf{x}},{\bf{x}}^{'})
=-G^{\sigma \sigma^{'}}({\bf{x}}, {\bf{x}}^{'})
 \psi^{\sigma'} ({\bf{x}}^{'}) 
\psi^{\sigma} ({\bf{x}})$.  
Here $\psi^{\sigma} ({\bf{x}})=F[n^{\sigma}({\bf{x}})]$ is a function of 
density $n^{\sigma}({\bf{x}})$, and 
$G^{\sigma \sigma^{'}}({\bf{x}}, {\bf{x}}^{'})$ is the interaction strength.
Assuming only nearest-neighbor interactions
for simplicity, and using 
$\psi^{\sigma}({\bf{x}})=n^{\sigma}({\bf{x}})$ with
$G^{\sigma \sigma}_{a}=0$ and $G^{\sigma \sigma'}_{a}\not=0$ for
$\sigma \not= \sigma'$, one obtains
\begin{equation}
\frac{d{\bf{p}}^{\sigma}}{dt}=
- n^{\sigma} ({\bf{x}}) \sum_{\sigma'} \sum_{a} 
G^{\sigma \sigma^{'}}_{a} n^{\sigma'} ({\bf{x}}+{\bf{e}}_{a}) {\bf{e}}_{a}
\end{equation} 
In a practical numerical study, one often assumes that 
$G^{\sigma \sigma'}_{a}$ is given by a constant $G$, and then varies 
the value of $G$
to model the surface tension strength. However,
one must be careful about the treatment of 
$G^{\sigma \sigma^{'}}_{a}$, and should use the values
that insure the Galilean-invariance of the 
macroscopic equation in three-dimensional space.
For the three-dimensional D3Q19 lattice model we use in the present study, 
after the  correct projection 
from 4D FCHC lattice \cite{re:ma96}, one obtains
\begin{equation}
\left\{
\begin{array}{lll}
G^{\sigma \sigma'}_{a}=G   , |{\bf{e}}_{a}|^{2}=1, \\
G^{\sigma \sigma'}_{a}=G/2 , |{\bf{e}}_{a}|^{2}=2, \\
G^{\sigma \sigma'}_{a}=0   , {\rm otherwise}.
\end{array}
\right.
\end{equation}

\section{Numerical Simulations and Discussion}
\subsection{Small Deformation Limit of Droplet Shear}
We now present our numerical study of 
the classical Taylor experiment, i.e. the deformation of 
single droplet under simple shear flow. The shear velocity is given by
\begin{equation}
{\bf{v}}=[\dot{\gamma} z,0,0]=[2Uz/L_{z},0,0]
\end{equation}
where $-L_{z}/2 \le z \le +L_{z}/2$, and $L_{z}$ 
is the distance between the top boundary plane 
which moves with velocity $+U$ and bottom boundary plane
which moves with velocity $-U$.  Taylor obtained 
a theoretical prediction for the dimensionless 
{\it small} deformation $D$.
With $\lambda=1$ and $\kappa=1$, where $Ca$ defined as before, we have
\begin{equation}
{\it Ca}=\frac{\dot{\gamma} R \nu \rho }{\beta}=
\frac{U}{L_{z}/2}\frac{R \nu \rho }{\beta}.
\end{equation}
Here $\nu$ and $\rho$ are the kinematic viscosity 
and the density of the fluid, respectively.
In our numerical simulation, we have chosen the 
densities $\rho = \rho_{1} = \rho_{2} = 0.3$ with $m^{1}=m^{2}=1$, and  
viscosities $\nu = \nu_{1} = \nu_{2} = (2\tau-1)/6$ with 
$\tau=\tau_{1}=\tau_{2}$=1.0. The initial radius $R$ of the drop was
set to 10.0. The system size was chosen to be 
$L_{x} \times L_{y} \times L_{z}$= 128 $\times$ 62 $\times$ 62.
We used parameter values $G_{12}=G_{21}$=0.5,
and $G_{11}=G_{22}$=0.0 for modeling the surface tension.
The actual value of surface tension $\beta$ was
determined by placing a
single droplet in the medium in the absence of shear flow, 
and using the Laplace law  $\delta p= 2\beta/R $ 
where $\delta p$ is the pressure difference inside and outside the
droplet and $R$ isthe radius of the droplet. Once we have all the numerical values 
($R, \nu, \rho, \beta, L_{z}$), the only 
free parameter left is the shear velocity $U$, which we can vary
to determine the Taylor number ${\it Ca}$.  The
initial condition for the simulation was taken to be a  single droplet
placed in the center of the computational volume with equal constant 
density both inside and outside the droplet. The initial macroscopic 
velocity field $\bf{u}$ 
was set to zero everywhere. In order to numerically
determine the ``major'' and ``minor'' axes $L$ and $B$,
we follow the standard technique used in classical mechanics
for calculating the rotational inertia $I_{ij}$. We first calculate
the symmetric matrix $A_{ij}$ defined as  
\begin{equation}
A_{ij}=\frac{\int\int\int \rho x_{i}x_{j} dx dy dz}
{\int\int\int \rho dx dy dz}
\end{equation} 
where the origin of the coordinates
is located at  the center of the initial droplet, and 
$x_{i}, i=1,2,3$ are the cartesian coordinates at a point in the
computational volume. After determining the
eigenvalues of the matrix $A_{ij}$,  we can then 
determine the values for $L$ and $B$. We made a series of runs of
the three dimensional LBM code for several small deformation-generating
shear flows.  For each such run we computed the values of $L$ and $B$ as 
indicated above. Figure 1 shows the 
comparison between our numerical results and the theoretical 
prediction for $D$ {\it vs} ${\it Ca}$. From the figure one can see
that the agreement between our numerical simulations with the LBM and
the theoretical result is excellent at small Taylor
number.  This agreement between our LBM-computed values and 
the theoretical value for small deformation is clear evidence 
that the numerical methods perform well at small Taylor number. This
validates the accuracy of the LBM plus SC model in this environment. 

\subsection{Larger Deformation to Breakup}

Given the successful comparison with the Taylor result in the small deformation
limit, we have confidence that the LBM plus the SC interface force model does
a credible job. We have performed several simulations at larger shear rates to
investigate the ability of the SC model to adequately track the deforming interface
to and beyond breakup.  Here we report the results of using the three dimensional
LBM code for the breakup. Our computational volume was 128 x 52 x 31.  The value of
the shear velocity for this was $U=0.5$.  The smaller value for $L_z$ results in
an effectively larger shear rate for the same value of $U$. This change results in
a qualitatively different range of the droplet deformation and breakup.  We emphasize that
the same values of the $G_{12}$, $G_{21}$,$G_{11}$, and $G_{22}$ used in the small
deformation limit were used for the breakup simulations.  The combination of LBM
and the SC model for the interface force between the two components proved to be a
robust combination once the initial work was done to determine the best range for the
$G_{12}$, $G_{21}$,$G_{11}$, and $G_{22}$ parameters. 

Figures 2 through 6 show snapshots of the evolution of the droplet under shear.
The initial spherical droplet was given a radius of 8.0.  The progression from
deformation to breakup is illustrated in the figures via the rendering of the
isosurface of the droplet in the surrounding volume of the other fluid. The
figures show just the fluid corresponding to the initial droplet.  A slice was taken
in a vertical plane through the center of the initial droplet and the results
rendered on a plane which is shown at the back side of the volume. Similarly, a second
slice was taken along the long axis and parallel to the bottom of the computational
volume  and projected onto the bottom of the volume's
bounding box.  These data show that there is considerable stretching and then pinching
breakup at the ends of the long, deformed droplet.  Subsequent to the breakup the shear
induces further stretching of the newly formed droplets. These in turn are eventually
significantly sheared to breakup themselves.  Figure 6 shows the droplets wrapping around
at the long ends of the volume due to our use of periodic boundary conditions at the ends
in the x-direction.

The generalization of these simulations to treat multiple droplet shear, breakup, and
coalescence is the next step in the simulation of these systems, and is under current
investigation.
 
\section{Conclusions}
In the present paper we have used the LBM along with the interface model
developed by Shan and Chen  \cite{re:sc93} 
for multiphase fluids to study the
single droplet deformation under simple shear flow. 
We have considered a three-dimensional two-fluid system with 
equal kinematic viscosities and densities. Our choice
of equal density and kinematic viscosity was made simply because of
the existence of the exact theoretical result against which we could compare
our numerical simulations of the deformation of the droplet.
We calculated the deformation $D$ and found that the numerical results
are in excellent agreement with Taylor's theoretical results.
We have also simulated the effect of larger shear rates and have successfully
evolved the sheared single droplet to breakup and beyond.
Thus this new LBM scheme is not limited to  small deformations and is of 
utility for the study 
of the dynamics deformation and breakup of many droplets systems.  It  could 
provide the capability to study the rheology of dispersed droplets and the 
dynamics of many droplets systems. Our results indicate that 
the LBM scheme is fully capable of prediciting the merger as well as
break-up \cite{re:xidun98} of many droplets systems.  Such studies are currently
underway. 
The results reported in this paper demonstrate that the LBM scheme which we have
utilized could be a useful tool  
for a wide range of industrial
problems including polymer molding processes and the rheology of many droplets systems.

One feature of the present method which needs further study centers around the
slightly non-robust character of the interface model. Namely, 
given a choice of the density ratio of the two fluids we have found the
choices of the $G_{12}$, $G_{21}$, $G_{11}$, and $G_{22}$ parameters characterizing
the strength of the interfacial forces which give a stable evolution to reside in 
a somewhat narrow range. Inside the range
the evolution is stable and produces qualitatively correct tracking of the breakup
process. Outside the range, the simulations eventually fail by producing negative
$f_i$ for one or both fluids thus inducing the halting of the computations.  It would be
useful to get a handle on how to extend the range of robustness of the simulations so
that a wide range of both the density ratio and the $G$ values can be used to increase 
the dynamic range of systems to which the methods can be applied. 
We are currently exploring the extent to which the use of the finite difference methods
with the extra freedom one gains by unlocking the velocity space from the position 
space lattice and requiring that the Courant limit be satisfied.
When used in conjunction with the SC interface model
it will be more stable and perhaps allow a wider
range of density ratios and $G$ values.  
We expect to report on our investigations into these 
matters in a future publication.

We thank Hudong Chen, Xiaowen Shan, Shiyi Chen, Nicos S. Martys, Xiaoyi He, and Li-Shi Luo for 
helpful and enjoyable discussions. 
This work was supported in part by NSF grant ASC-9418357 for the Pharoh 
MetaCenter Regional Alliance and by the PRF under contract No. 33160-GB9.  
The simulations were performed on the SGI PowerChallenge  at the Ohio
Supercomputer Center.  We thank Ken Flurchick of the Ohio Supercomputer
Center for
his expert assistance with the visualization of the droplet shear and
breakup.

\vfill\eject
\begin{figure}[bth]
\begin{center}
\makebox[4in]{\epsfig{file=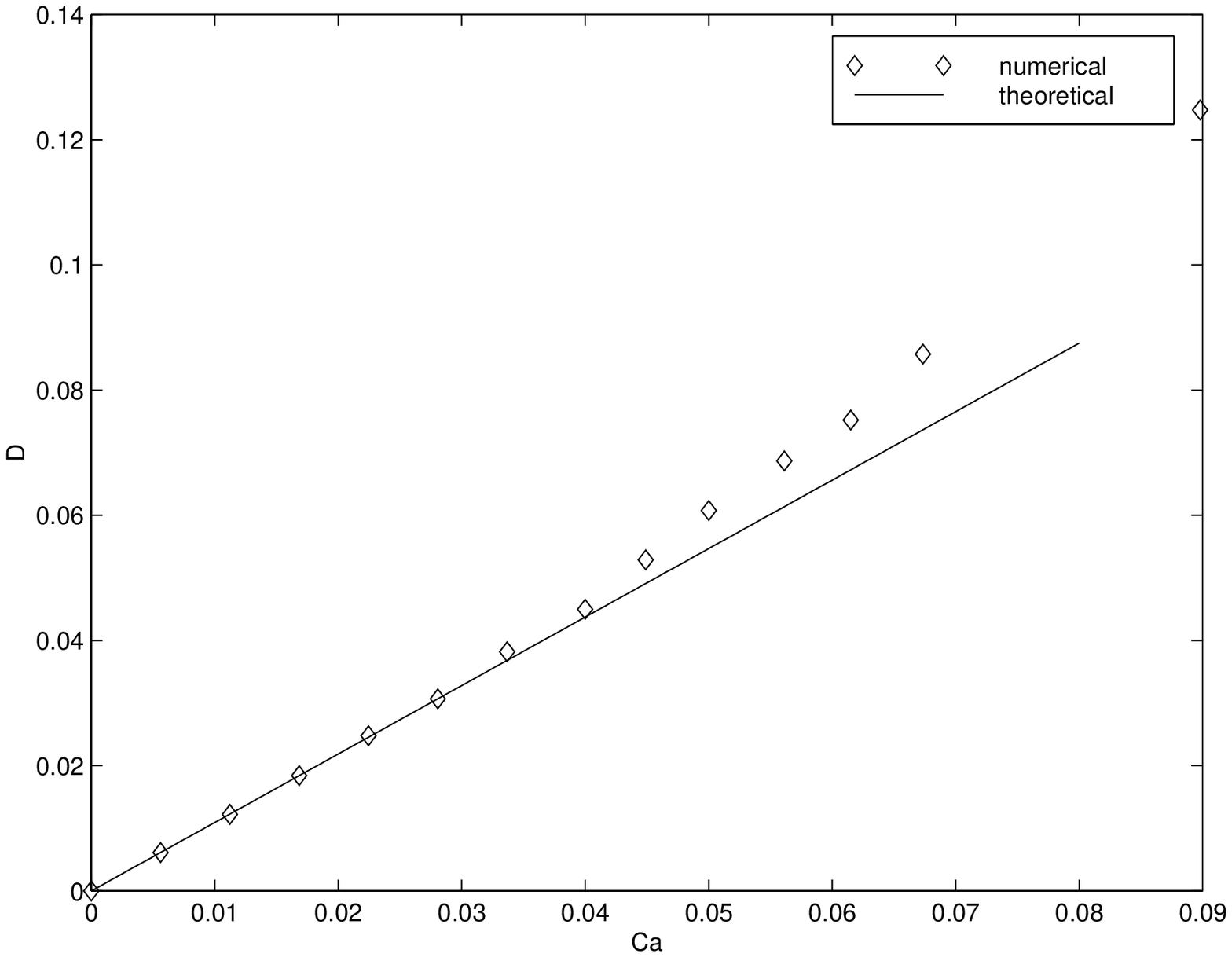,width=6.5in}}
\caption{Plot of deformation vs Taylor number in the small deformation limit.
Shown also is the theoretical result of Taylor.}
\label{fig1}
\end{center}
\end{figure}
\begin{figure}[bth]
\begin{center}
\makebox[4in]{\epsfig{file=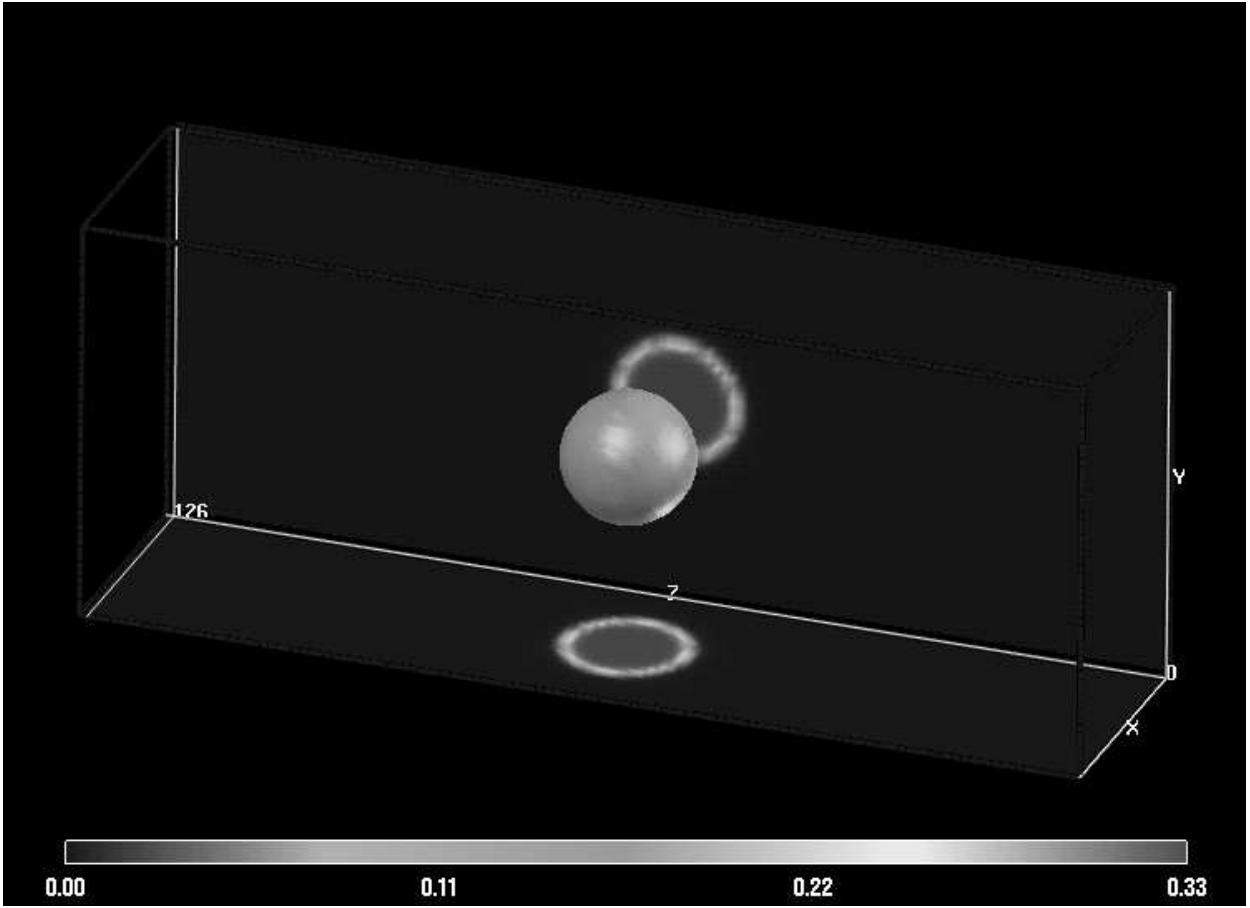,width=6.5in}}
\caption{Initial Droplet of radius 8.0. The legend shows a grey scale
colormap of the density of the sheared droplet. }
\label{fig2}
\end{center}
\end{figure}
\begin{figure}[bth]
\begin{center}
\makebox[4in]{\epsfig{file=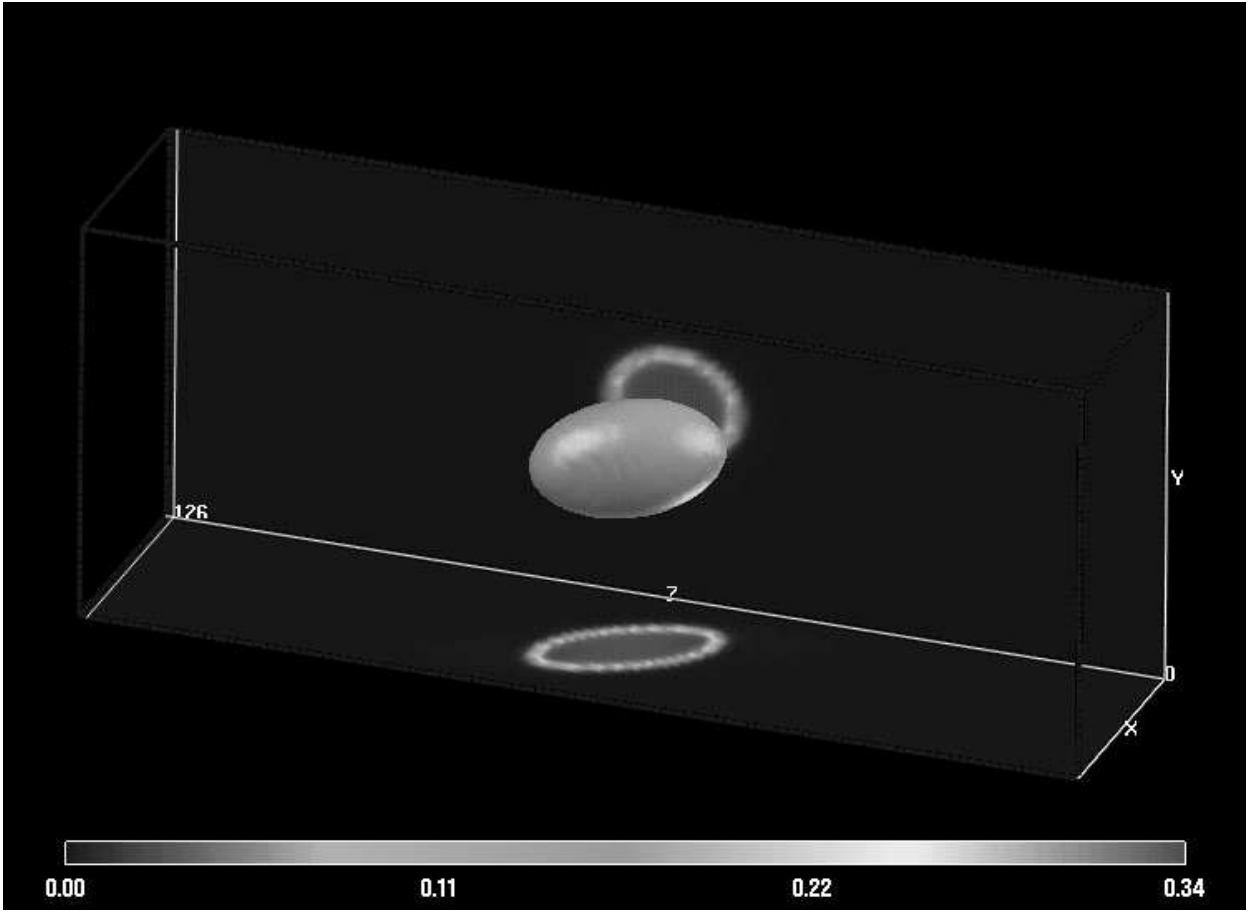,width=6.5in}}
\caption{Deformed droplet at 400 cycles }
\label{fig3}
\end{center}
\end{figure}
\begin{figure}[bth]
\begin{center}
\makebox[4in]{\epsfig{file=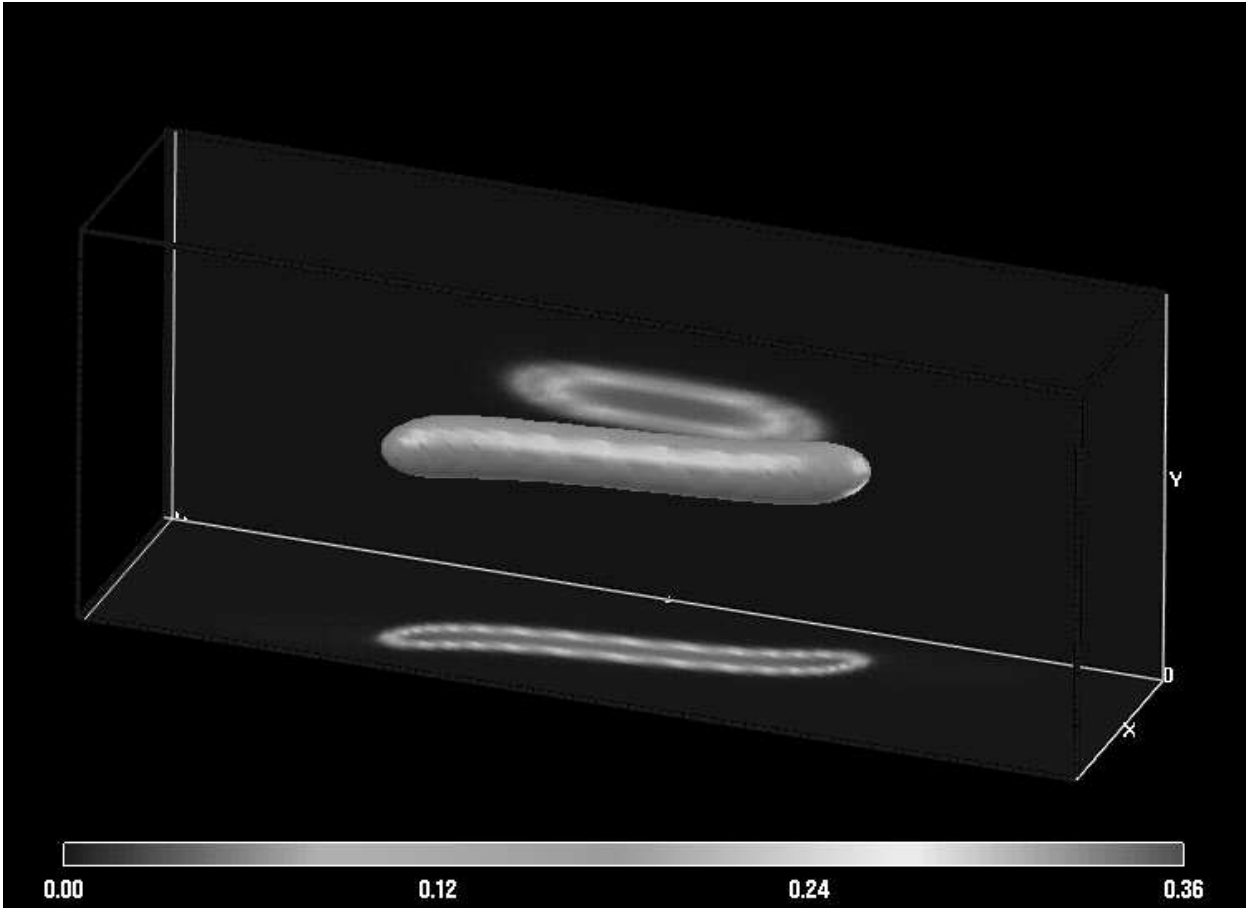,width=6.5in}}
\caption{Sheared droplet at 800 cycles }
\label{fig4}
\end{center}
\end{figure}
\begin{figure}[bth]
\begin{center}
\makebox[4in]{\epsfig{file=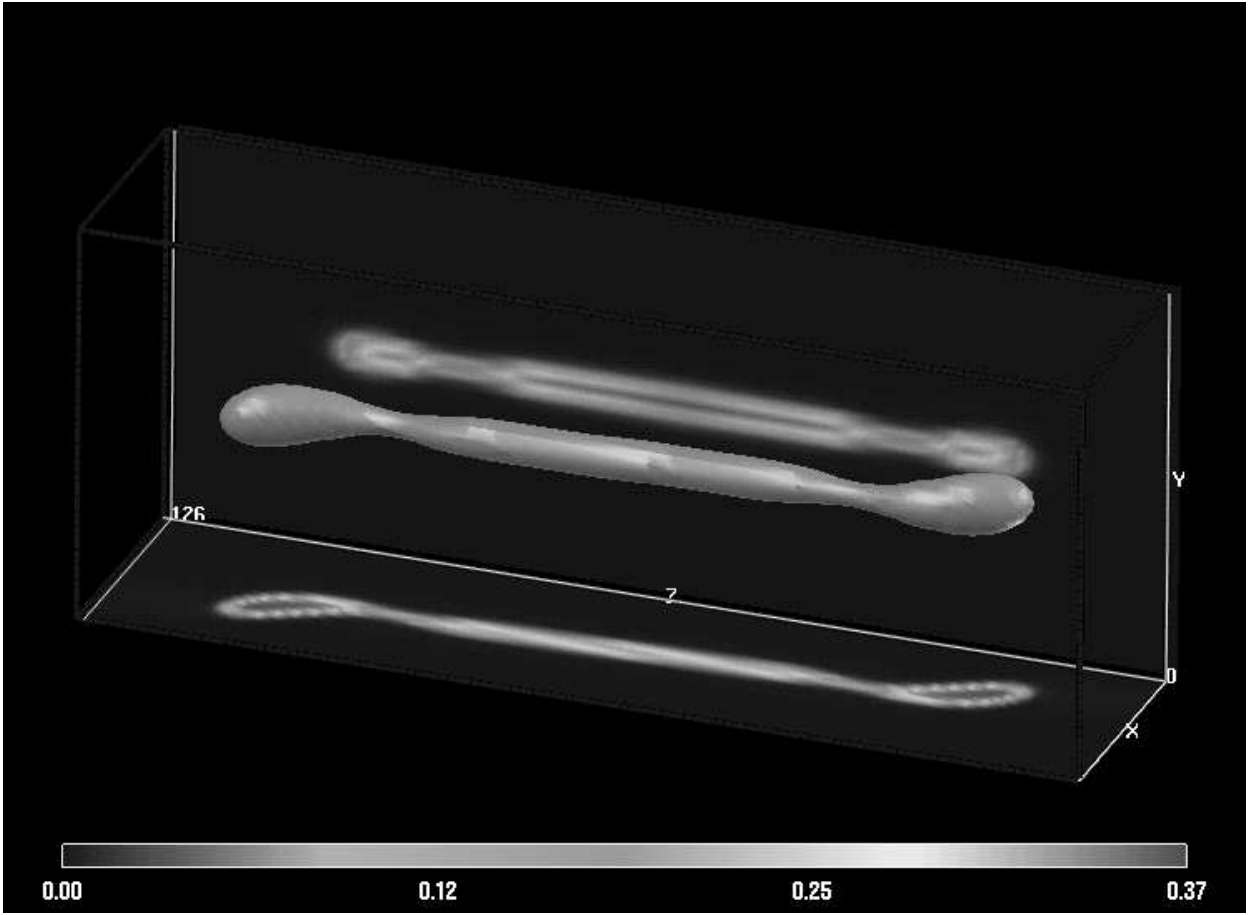,width=6.5in}}
\caption{Droplet at 1200 cycles just before pinched breakup at its ends}
\label{fig5}
\end{center}
\end{figure}
\begin{figure}[bth]
\begin{center}
\makebox[4in]{\epsfig{file=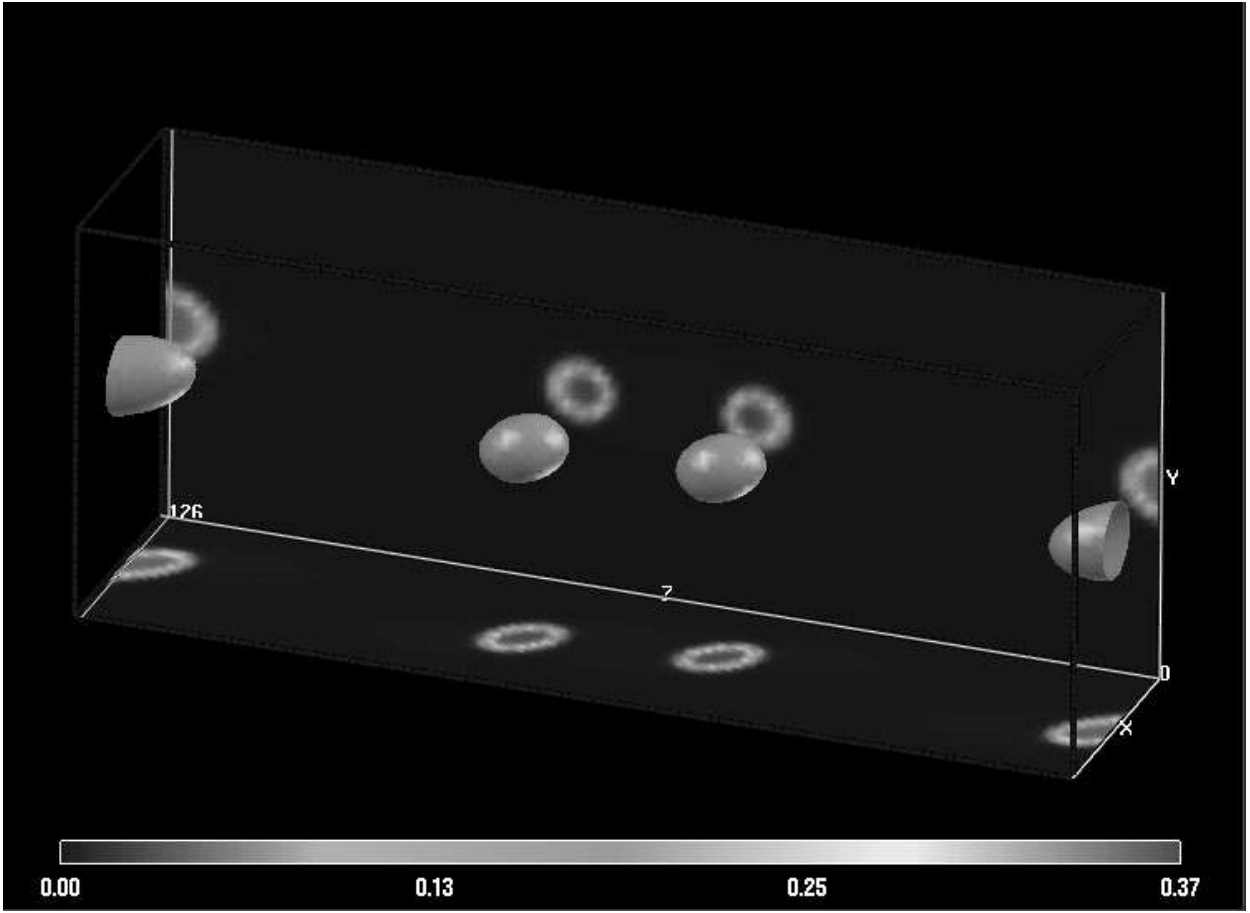,width=6.5in}}
\caption{Multiple droplets at 1600 cycles }
\label{fig6}
\end{center}
\end{figure}

\end{document}